# Assessing national strengths and weaknesses in research fields[1]

*Giovanni Abramo*[*]
  Laboratory for Studies of Research and Technology Transfer
  Institute for System Analysis and Computer Science (IASI-CNR)
  National Research Council of Italy
  Viale Manzoni 30, 00185 Rome - ITALY
  giovanni.abramo@uniroma2.it

*Ciriaco Andrea D'Angelo*
  Department of Engineering and Management
  University of Rome "Tor Vergata"
  Via del Politecnico 1, 00133 Rome - ITALY
  dangelo@dii.uniroma2.it

**Abstract**

National policies aimed at fostering the effectiveness of scientific systems should be based on reliable strategic analysis identifying strengths and weaknesses at field level. Approaches and indicators thus far proposed in the literature have not been completely satisfactory, since they fail to distinguish the effect of the size of production factors from that of their quality, particularly the quality of labor. The current work proposes an innovative "input-oriented" approach, which permits: i) estimation of national research performance in a field and comparison to that of other nations, independent of the size of their respective research staffs; and, for fields of comparable intensity of publication, ii) identification of the strong and weak research fields within a national research system on the basis of international comparison. In reference to the second objective, the proposed approach is applied to the Italian case, through the analysis of the 2006-2010 scientific production of the Italian academic system, in the 200 research fields where bibliometric analysis is meaningful.

**Acknowledgement**

*We wish to thank two anonymous referees for their constructive and valuable suggestions.*

---

# 1. Introduction

Research activity conducted in universities and research institutions is a crucial driver for innovation, competitiveness and the socio-economic progress of nations (Griliches, 1998; Henderson et al., 1998; Mansfield, 1995; Rosenberg and Nelson, 1994; Adams, 1990). Universities and research institutions provide the life-blood of the knowledge-based economy, through the formation of human capital, the advancement of knowledge in the different fields of science, the development of new technologies and applications, and in licensing and creation of high-tech spinoff firms (Etzkowitz et al., 2000; Martin et al., 1996; Mansfield, 1991). Awareness of these roles underlies the growing numbers of convinced supporters for policies aimed at reinforcing higher education and research systems, through investments and added funding programs. Many governments have remained faithful to such strategies in spite of the budgetary effects of the global economic crisis, as seen in recent years. However the limitations on public resources, coupled with simultaneous increases in social needs, have forced governments to pay more attention to the efficiency and effectiveness of their interventions. In the research sphere this translates into more attentive and rational allocation of resources. Policy-makers ideally seek effectiveness through identification of the scientific sectors with the highest potential of socio-economic returns, and efficiency through award of competitive funding to the most productive researchers and research institutions.

In cases where national research funds are primarily allocated through calls for proposals, it is possible to pursue both objectives at once. Effectiveness is sought through the identification of so-called strategic sectors and the division of available funds among different sectors according to degrees of priority. Efficiency is pursued through allocation of resources to the best projects in each sector. However in nations where the largest share of funding is allocated directly to the overall research institutions, with greater or lesser levels of competition, the options for strategic allocation of resources are limited and the effectiveness of interventions is jeopardized. Many nations offering direct institutional funding have thus determined to adopt national research assessment exercises[2], and allocate resources to their institutions on the basis of the results. This approach is indeed functional for pursuing efficiency in public interventions, but not necessarily for effectiveness: regardless of any alteration in the funds awarded, the top research institutions could, in part or in whole, conduct their research in sectors of little or no strategic priority; meanwhile, the worst research institutions could be the ones dealing in sectors of highest priority.

In the formulation of the national research assessment exercises, especially those using bibliometric indicators, it is actually possible to observe choices with undesirable strategic implications. This is the case with the most recent Italian national evaluation exercise, the VQR (Research Quality Evaluation), completed in July 2013. Here the quality of the hard sciences research products submitted by institutions was evaluated by number of citations (and in some cases impact factor) standardized with respect to an international benchmark, rather than a national one. In this manner the institutions with greater concentration in research fields where Italy is a follower or late follower are penalized compared to those with a greater concentration of research in fields where Italy is at the frontier. This penalization not only appears unjust, but could in fact be counterproductive for the national system, in the case of strategic motivations for fostering catch-up research in the "follower" fields rather than the frontier research in "world-class" fields.

---

[2] According to a recent study by Hicks (2012), there are currently 15 nations (China, Australia, New Zealand, 12 EU countries) that conduct regular comparative performance evaluations of universities and link the results to public financing. The shares of overall public funding and the criteria for assigning funds vary from nation to nation.

In any case, prior to the formulation of any policy intended to improve the effectiveness of a national research system, governments would obviously be well advised to conduct strategic analysis to identify the strong and weak research fields of their respective national systems. However the analytical methods for this task as thus far proposed in the literature present a number of limitations, and it is to this theme that the authors now direct their attention.

The next two sections of our paper provide a brief review of the literature on measuring the scientific standing of nations and an examination of the methodological shortcomings of the current approaches. Section 4 presents the methodology proposed by the current authors and the bibliometric dataset used to test it, referring to the Italian academic system. Section 5 presents the results of the analyses conducted at the aggregate level of the disciplines, and Section 6 provides a deeper investigation at the level of scientific fields. The concluding section summarizes the work, indicates potential applications for the proposed method, and offers the authors' suggestions for future directions on the theme.

## 2. Measuring scientific standing: literature review

Defining, measuring and comparing the "scientific standing" of institutions or nations in the different scientific fields is a difficult and challenging responsibility for scholars in the field, given the multidimensional and highly complex character of the tasks (Hauser and Zettelmeyer, 1997; Werner and Souder, 1997). For Tijssen (2003), scientific standing has a comparative character, implying "surpassing something or someone in quality", and for him the most important drivers are: i) the creation of new scientific and technical knowledge; ii) its transmission to user communities; iii) the commercial exploitation of that knowledge.

In fact there is no unanimous opinion on the meaning of "research standing", much less on the relative indicators for its measure. However there is a certain agreement on the fact that standing has a strong link with "research quality" and "research impact", even though some scholars hold that impact is a part of research quality (Boaz and Ashby, 2003; OECD, 1997), the other parts being importance and accuracy of research (Martin and Irvine, 1983), while others hold that quality and impact are two different elements of research standing (Grant et al., 2010).

Recent progress in techniques of bibliometric measurement has certainly provided a significant push to studies on the measurement of research standing, conducted both at the level of institutions and national systems. May (1997) provides a first definition of research standing: "For many purposes, most notably overall advance in our understanding of nature, it is total output that matters. For other purposes - for example, in producing trained people or for underpinning industrial advances - output relative to country size is more relevant". He then measured the relative international standing of 15 countries in science, medicine and engineering, by their shares of ISI-indexed publications and citations as well as by citations per unit of spending, over a 14-year period. May also calculated the comparative advantage of countries in each of 20 disciplines, by the fraction of a country's citations in a discipline relative to the world's fraction. A year later, Adams (1998) presented a study sponsored by the Higher Education Funding Council of England, aimed at identifying England's relative strengths and weaknesses in performance when comparing between fields. Impact measures for England in 47 disciplines were compared over a 9-year period to those of six other nations, against a world baseline. King (2004) updated May's original 1997 work to 2002, covering a 10-year period. The new study increased the number of nations analyzed (31), provided a longitudinal analysis over two five-year periods, added further indicators (top 1% highly cited articles; average citations per paper), provided for normalization of citations to the mean for each field, and took account of year of publication, thus providing aggregate measures of the overall research standing of each country. Outputs and outcomes were also normalized to inputs (researchers, expenditures, GDP) at aggregate level.

In the past decade, studies concerning the relative standing of nations have increasingly tended to focus on excellence, and highly-cited articles (HCAs) have become the proxy for identifying it. This choice is based on certain assumptions, which have been broadly accepted in the literature: i) in the *hard sciences*, the prevalent form of codification of research output is publication in peer reviewed journals, so that it is assumed that excellent results are observable in the form of excellent publications; ii) the excellence of a publication is demonstrated by its placement in the high extremes of the scale of value shared by the international scientific community of the specific discipline; iii) the value of a publication is understood as its impact on scientific advancement, and as proxy of this impact bibliometricians adopt the number of citations for the publication itself. Tijssen et al., (2002) proposed a citation-based "systems approach" for analyzing the various institutional and cognitive dimensions of scientific excellence within national research systems. The methodology, which covers several aggregate levels, focuses on HCAs in the international journal literature. Pislyakov and Shukshina (2012) take HCAs as a proxy for "excellence" and co-authored papers as a measure of collaborative efforts, to discover Russian "centers of excellence" and explore patterns of their collaboration with each other and with foreign partners. Bornmann and Leydesdorff (2011) have developed new approaches for the spatial visualization of concentrations of HCAs using overlays to Google Maps. Similarly, for mapping field-specific centers of excellence around the world, Bornmann et al. (2011) used bibliometric data to identify cities where highly-cited papers were published. Finally for some years, research groups as the CWTS of the University of Leiden[3], the SCImago group[4] and others have published on-line country ranking, using bibliometric indicators such as total number of articles, average normalized citations per article and HCAs.

---

[3] http://www.leidenranking.com/ranking, last accessed on June 23, 2014.
[4] http://www.scimagojr.com/countryrank.php, last accessed on June 23, 2014.

## 3. Shortcomings of current approaches

The approach and the indicators applied in the literature to date, while useful for some purposes, do not prove completely satisfactory. To the current authors, the greatest concern over the approach is that it does not succeed in separating the effect of size (labor and capital) from the effect of the quality of the production factors, particularly labor. In all the literature, the USA invariably ranks at the top for such indicators as number and share of both publications and HCAs, in all scientific sectors. But does the USA's performance depend on the fact that it has larger research expenditures than every other individual nation, or is it truly because American scientists are better than the others? The only responses to this question, attempted by normalizing outputs and outcomes to inputs, have dealt with the data at the aggregate level and have not been terribly effective (Abramo and D'Angelo, 2007). In fact while most scholars now typically normalize the observed output data, accounting for the field and year of publication, the data on input are not correspondingly divided according to the fields of allocation, since the practitioners lack data on the numbers of researchers and the expenditures per field in the individual countries under comparison. The latest attempt by Bornmann et al. (2014) cannot do any better than normalizing output indicators by GDP per capita.

Some of the indicators used for output also leave much to be desired. The simple count of publications can be misleading as an indicator of scientific strength: one nation could have many more publications than another, but of markedly lesser quality (impact). The value of average citations per publication, whether normalized or not, is also inconsistent. For example a nation with 1,000 publications in a field, each one with 10 citations, would rank higher than a nation with 10,000 publications, of which 9,999 have 10 citations but the last one a mere nine. In this work we propose an approach that is different from the output-oriented methodologies repeatedly found in the literature. Instead of beginning from output or outcome, we depart from input, with the objective of controlling for the effects of size. Further, we explicitly assert that one country is more productive than another one in a given field if its researchers are more productive than the others', independent of their number. Analogously, within a single country, a field is stronger than another one if in the relative international comparison of fields it ranks higher than the other one. The ideal indicator for measuring such comparisons is research productivity at the field level, but to calculate such an indicator requires knowledge of at least the number of researchers per field, as well as their output. Unfortunately the data on numbers of scientists in the different nations do not include classification by field of research. Since we cannot measure the research productivity of each nation, and our objective is not to rank the world performance in the individual fields, but rather the national one, we have devised an alternative method. Our approach, which is input oriented, provides for the classification of individual researchers by scientific field, as well as the bibliometric evaluation of their performance in international terms. The methodology is described below.

## 4. Methodology: requirements, assumptions and bibliometric dataset

In applying our methodology we take advantage of a characteristic that seems unique to the Italian research system, in which each academic is officially classified as belonging to a single specifically-defined research field, called a "Scientific Disciplinary Sector" (SDS). The national faculty system is composed of 370 SDSs, grouped into 14 "University Disciplinary Areas" (UDAs). In other countries lacking a similar system it would still be possible to attempt to classify researchers by identifying the prevalent subject category of their publications. With such a classification for a range of countries and an exhaustive bibliometric dataset of the scientific production of the individual scientists, it would be possible to compare a proxy of labor productivity in each field over specific periods, where labor productivity is understood as the average normalized impact per researcher (not publication) in the different countries. Knowing that such classifications and datasets have not yet been developed, we devise a different route to pursue the objective of this work, specifically: "how are we to identify the strong and weak research fields of a country, in a manner that can inform research policies and initiatives?". We will resort to the HCAs, which we define here as those publications that place in the top 1% ($HCAs_{1\%}$) or 5% ($HCAs_{5\%}$) of the world citation rankings for WoS-indexed publications[5] of the same year and subject category[6]. For fields of comparable intensity of publication, we will qualify one field as stronger than another if the quotient of researchers publishing HCAs to total in the field is higher. The underlying rationale is that the higher the concentration of researchers in a field who can produce highly-cited publications, meaning being able to advance the frontier of knowledge in that field, the relatively stronger is that field in the country. If the intensity of publication is notably different among fields, then fields with higher intensity would be favored, because the probability of having an article among the highly cited ones increases with the number of articles produced, which depends on both the quality of a scientist and the average intensity of publication in the field. Therefore there is a bias in favor of fields in which researchers publish a lot. To overcome this problem one should know the average intensity of publication in each field, which we lack at the moment. This bias may be reduced a bit by using fractional counting, instead of full counting of publications. A minor bias in the method also occurs in favor of those fields where the ratio of domestic scientists to world ones is higher. In fact, all other equals, the probability of having top scientists in a field decreases as the ratio decreases. To overcome this other bias is a formidable task as well, since the classification of scientists by field is lacking at world level. The findings of the methodology applied to the Italian research system should be interpreted having in mind the above limitations.

Drawing on the classification of all Italian professors[7] in their research fields, we proceed in the following manner: i) beginning from the raw data of the Web of Science (WoS) over the period 2006-2010, and applying an algorithm for reconciliation of the author's affiliation and disambiguation of their precise identity, we attribute each publication to the university scientist that produced it[8] (D'Angelo et al., 2011); ii) we

---

[5] For publications in multi-category journals, we consider the percentile for the most favorable category.

[6] It would also be possible to choose thresholds other than 1% and 5%. Glänzel and Schubert (1988) provide further discussion on this subject.

[7] Unfortunately we cannot include in our analysis scientists from research institutions because they lack SDS classification.

[8] Our author disambiguation approach follows a three-step process: database integration, mapping generation, and filtering. First, information from a database of all Italian professors maintained by the Italian ministry of university and research is integrated into the bibliometric database. As a result, a reference list of author identities and their attributes is added to the original database. Second, a mapping algorithm links each author of an article to all the possible author identities from the reference list. Finally, different data-driven heuristics are used to filter out as many false positives as possible. The

identify all HCAs and the number of researchers with at least one HCA over the period, who we will call "top scientists" (TSs); iii) we measure the ratio of TSs in a field to the total number of researchers in that field. With this approach, for fields of comparable intensity of publication, we succeed in distinguishing those fields in which the Italian research system is relatively strong from those where it is weak, on the basis of the percentage of TSs at the international level, controlled for field size. Since the intensity of joint research work varies across fields (Abramo et al., 2013), to control for both the high intensity of publication and co-authorships in some fields, where the number of TSs could also result as relatively higher, we repeat the exercise using fractional counting of HCAs and adopt the convention of defining a TS as an academic with a total fractional counting of HCAs that exceeds a certain threshold.

For reasons of robustness, the study is limited to those fields where bibliometric analysis can be considered significant. Thus the field of observation is limited to the SDSs where over the five years examined, at least 50% of Italian professors achieved at least one publication indexed in WoS: this results as 200 SDSs, belonging to 11 UDAs[9]. The 200 fields included roughly 39,525 professors[10] that were on faculty for at least three years over the 2006-2010 period[11], who produced almost 200,000 WoS-listed publications. Table 1 presents the distribution of publications per UDA.

*Table 1: Dataset for the analysis, per UDA (data 2006-2010)*

| UDA | No. of SDSs | Research staff | Publications* | HCAs(1%) | HCAs(5%) |
|---|---|---|---|---|---|
| Mathematics and computer sciences | 9 | 3,337 | 15,755 | 138 (0.9%) | 680 (4.3%) |
| Physics | 8 | 2,617 | 23,511 | 322 (1.4%) | 1,476 (6.3%) |
| Chemistry | 12 | 3,312 | 25,494 | 324 (1.3%) | 1,542 (6.0%) |
| Earth sciences | 12 | 1,272 | 5,215 | 67 (1.3%) | 309 (5.9%) |
| Biology | 19 | 5,339 | 30,977 | 428 (1.4%) | 1,829 (5.9%) |
| Medicine | 49 | 11,309 | 62,852 | 948 (1.5%) | 3,904 (6.2%) |
| Agricultural and veterinary sciences | 29 | 2,930 | 11,643 | 123 (1.1%) | 576 (4.9%) |
| Civil engineering | 9 | 1,575 | 5,309 | 45 (0.8%) | 190 (3.6%) |
| Industrial and information engineering | 41 | 5,159 | 36,947 | 215 (0.6%) | 1,205 (3.3%) |
| Pedagogy and psychology | 7 | 934 | 3,338 | 32 (1.0%) | 180 (5.4%) |
| Economics and statistics | 5 | 1,741 | 3,437 | 32 (0.9%) | 134 (3.9%) |
| Total | 200 | 39,525 | 196,857† | 2,279† (1.2%) | 10,372† (5.3%) |

*\* Publications over 2006-2010 authored by at least one Italian professor from the UDA (considering only professors with at least three years on faculty over the period).*
*† The total value is different than the sum of values per row due to multiple counting of publications co-authored by Italian professors in different UDAs.*

The data show the predominance of Medicine concerning all the dimensions reported. Researchers in this discipline alone represent 28.6% of the total dataset, producing 31.9% of the publications, with 42% of total HCAs(1%) and 38% of HCAs(5%).

---

result of the last step is a robust mapping between author instances and author identities with a minimum number of false positives and a negligible number of false negatives. The harmonic mean of precision and recall (F-measure) of authorships disambiguated by our algorithm is around 96% (2% margin of error, 98% confidence interval).

[9] The complete list is accessible at www.disp.uniroma2.it/laboratoriortt/TESTI/Indicators/ssd4.html, last accessed on June 23, 2014

[10] The dataset of Italian professors is extracted from a database maintained by the Ministry of Education, Universities and Research (http://cercauniversita.cineca.it/php5/docenti/cerca.php, last accessed on June 23, 2014).

[11] See Abramo et al., 2012, for details about this choice.

Concerning the HCAs$_{1\%}$, we observe that they represent 1.2% of total publications in the dataset, with peaks in Medicine (1.5%) and Physics and Biology (both 1.4%). Considering instead the HCAs$_{5\%}$ (which overall represent 5.3% of total publications), the UDA with highest incidence is Physics (6.3%), followed by Medicine (6.2%) and Chemistry (6.0%).

## 5. Analysis at aggregate level (UDA)

As seen in Table 1, the dataset consists of 39,525 Italian professors (assistant, associate and full): 8.1% of these (3,195) appear at least once in the bylines for the 2,279 HCAs$_{(1\%)}$ recognized for the period 2006-2010 (Table 2).

The highest incidence occurs in the Physics UDA, which has an overall Italian research staff of 2,617 professors and 620 TSs$_{(1\%)}$ (23.7% of total faculty). Chemistry is next but substantially behind, with 11.3% of total research staff achieving at least one HCA over the period examined. Immediately following are Medicine and Biology, with incidence of TSs$_{(1\%)}$ below 10%, respectively at 9.4% and 7.8%. Last on the list are Pedagogy and psychology and Mathematics and computer sciences (both at 3.6%), Civil engineering (2.6%) and Economics and statistics (2.4%).

The particularly high incidence of TSs$_{(1\%)}$ in Physics is to some extent clearly due to the specific research collaboration behavior and the high intensity of publication in the discipline. Especially in the fields of particle and high-energy physics, research is often conducted through so-called "grand experiments". The results typically have high scientific impact and are credited to a large part of the research staff of the partner organizations. Research results are then codified in publications with hundreds or even thousands of co-authors. To control for the effects of such multi-authored and high number of publications we repeat the analysis with a fractional counting approach. We assume that each author, for each publication, is recognized for a contribution equal to the reciprocal of the number of co-authors. We then also assume that the TSs$_{(1\%)}$ are identifiable as the professors with a total fractional output (i.e. sum of contributions relative to each authored publication) equal to at least 0.1[12].

*Table 2: Italian professors authoring HCAs$_{(1\%)}$ per UDA ("full authorship" counting method; data 2006-2010)*

| UDA | Research staff | TSs$_{(1\%)}$ | Incidence (%) | Rank |
|---|---|---|---|---|
| Physics | 2,617 | 620 | 23.7 | 1 |
| Chemistry | 3,312 | 373 | 11.3 | 2 |
| Medicine | 11,309 | 1,061 | 9.4 | 3 |
| Biology | 5,339 | 416 | 7.8 | 4 |
| Earth sciences | 1,272 | 71 | 5.6 | 5 |
| Industrial and information engineering | 5,159 | 272 | 5.3 | 6 |
| Agricultural and veterinary sciences | 2,930 | 145 | 4.9 | 7 |
| Pedagogy and psychology | 934 | 34 | 3.6 | 8 |
| Mathematics and computer sciences | 3,337 | 121 | 3.6 | 9 |
| Civil engineering | 1,575 | 41 | 2.6 | 10 |
| Economics and statistics | 1,741 | 41 | 2.4 | 11 |
| Total | 39,525 | 3,195 | 8.1 | |

---

[12] It would also be possible to choose thresholds other than 0.1. We have conducted a sensitivity analysis for the Italian case: a higher threshold would noticeably reduce the number of TSs; a lower one would not significantly impact the ranks.

Table 3 presents the results of the new analysis: the total number of TSs$_{(1\%)}$ now drops to 1,918, or 4.9% of total Italian faculty. The differences observed in the individual UDAs are variable. Physics, which under "full-author" counting had the greatest incidence of TSs$_{(1\%)}$, now shows a number of TSs$_{(1\%)}$ corresponding to 6.3% of total research staff in the UDA, and is outdone by Chemistry, which itself drops to 9.8% from the previous 11.3%. The effect of fractionalization on the individual outputs of the other UDAs does not have particularly noticeable effects: the rank for incidence of TSs$_{(1\%)}$ remains substantially unvaried compared to under "full-authorship", and is particularly so for the four UDAs at the bottom. This indicates that, apart from the specific case of Physics[13], in disciplines of comparable intensity of publication, the methodology is sufficiently robust, in the sense of being quite free of effects that the different intensities of research collaboration could have on the chosen indicator.

*Table 3: Percentage of Italian professors authoring HCAs$_{(1\%)}$ per UDA (fractional counting method; data 2006-2010)*

| UDA | Research staff | TSs$_{(1\%)}$ | Incidence(%) | Rank |
|---|---|---|---|---|
| Physics | 2,617 | 166 | 6.3 | 2 |
| Chemistry | 3,312 | 325 | 9.8 | 1 |
| Medicine | 11,309 | 583 | 5.2 | 3 |
| Biology | 5,339 | 270 | 5.1 | 4 |
| Earth sciences | 1,272 | 43 | 3.4 | 7 |
| Industrial and information engineering | 5,159 | 228 | 4.4 | 5 |
| Agricultural and veterinary sciences | 2,930 | 119 | 4.1 | 6 |
| Pedagogy and psychology | 934 | 29 | 3.1 | 8 |
| Mathematics and computer sciences | 3,337 | 94 | 2.8 | 9 |
| Civil engineering | 1,575 | 36 | 2.3 | 10 |
| Economics and statistics | 1,741 | 25 | 1.4 | 11 |
| Total | 39,525 | 1,918 | 4.9 | |

We now ask whether the indicator is sufficiently sensitive to the threshold imposed for identifying the HCAs. For this, Table 4 presents the analysis again, but now based on the dataset of the top 5% of publications rather than the top 1%.

*Table 4: Italian professors authoring HCAs$_{(5\%)}$ per UDA (full counting method, data 2006-2010)*

| UDA | Research staff | TSs$_{(5\%)}$ | Incidence (%) | Rank |
|---|---|---|---|---|
| Physics | 2,617 | 1,162 | 44.4 | 1 |
| Chemistry | 3,312 | 1,128 | 34.1 | 2 |
| Medicine | 11,309 | 3,044 | 26.9 | 3 |
| Biology | 5,339 | 1,360 | 25.5 | 4 |
| Earth sciences | 1,272 | 268 | 21.1 | 5 |
| Agricultural and veterinary sciences | 2,930 | 593 | 20.2 | 6 |
| Industrial and information engineering | 5,159 | 973 | 18.9 | 7 |
| Mathematics and computer sciences | 3,337 | 477 | 14.3 | 8 |
| Pedagogy and psychology | 934 | 128 | 13.7 | 9 |
| Civil engineering | 1,575 | 158 | 10.0 | 10 |
| Economics and statistics | 1,741 | 110 | 6.3 | 11 |
| Total | 39,525 | 9,401 | 23.8 | |

Obviously the incidence of TSs$_{(5\%)}$ out of total research staff now increases: at the general level of Italian professors, 23.8% authored at least one HCA $_{(5\%)}$ over the 2006-2010 period. Maximum incidence is again seen in Physics, with 44.4% of faculty, followed as before by Chemistry (34.1%), Medicine (26.9%) and Biology (25.5%). In the ranking by UDA, the only variations are in the central area of the list, where there are exchanges of positions between Agricultural and veterinary sciences and Industrial and information engineering, as well as between Pedagogy and psychology and Mathematics and computer sciences. Thus in general, changing the threshold for identification of HCAs does not have significant impact on the results of the analyses.

---

[13] In this UDA the production function for new knowledge typically involves very extensive collaborations to arrive at results of highest excellence. Here, Italian researchers evidently achieve notable success.

## 6. Strengths and weaknesses at field level (SDS)

To identify the strong and weak points in a national research system it is obviously necessary to inquire at a greater level of detail than simply comparing the major disciplinary areas. For the Italian case, we offer the example of the analysis of the Earth sciences SDSs, as seen in Table 5.

In this discipline, the field that registers the maximum incidence of $TSs_{(1\%)}$, is GEO/03 (10.7%), followed by GEO/10 (10.1%) and GEO/12 (8.7%). Last on the list is GEO/05, which is the second largest field of the UDA in terms of total research staff, with 165 professors, of whom none authored an $HCA_{(1\%)}$ over the period examined. The data in the last column of Table 5 indicate the position of Earth sciences SDSs out of the total 200 investigated in all UDAs, in percentile terms (100 the best, 0 the worst): GEO/03, which leads the ranking for the UDA, places in 81$^{st}$ percentile in overall national ranking. Of the other 11 SDSs, seven place above median, from 78$^{th}$ to 52$^{nd}$ rank. The other four are all below median in general rank.

Repeating the analysis for all 200 of the SDSs under observation we observe that in 27 of these there are no TSs over the five years examined (Table 6). More specifically, this occurs for one SDS in Earth sciences (out of 12 in the UDA), three of 49 in Medicine, six of 29 in Agricultural and veterinary sciences, two out of nine in Civil engineering, 14 of 41 in Industrial and information engineering and one out of seven SDSs in Pedagogy and psychology. Among the SDSs with no TSs, six have a national research staff of over 100 faculty members.

*Table 5: Italian professors authoring HCAs$_{(1\%)}$ in the SDS of Earth sciences (full counting method; data 2006-2010)*

| SDS | Research staff | TSs$_{(1\%)}$ | Incidence (%) | Percentile rank (over 200 SDSs) |
|---|---|---|---|---|
| GEO/03-Structural Geology | 103 | 11 | 10.7 | 81 |
| GEO/10-Geophysics of Solid Earth | 79 | 8 | 10.1 | 78 |
| GEO/12-Oceanography and Atmospheric Physics | 23 | 2 | 8.7 | 72 |
| GEO/07-Petrology and Petrography | 113 | 9 | 8.0 | 66 |
| GEO/01-Palaeontology and Palaeoecology | 120 | 8 | 6.7 | 57 |
| GEO/08-Geochemistry and Volcanology | 97 | 6 | 6.2 | 56 |
| GEO/02-Stratigraphic and Sedimentological Geology | 192 | 11 | 5.7 | 54 |
| GEO/06-Mineralogy | 109 | 6 | 5.5 | 52 |
| GEO/04-Physical Geography and Geomorphology | 138 | 6 | 4.3 | 43 |
| GEO/09-Mineral Geological Resources and Mineralogic and Petrographic Applications | 82 | 3 | 3.7 | 34 |
| GEO/11-Applied Geophysics | 51 | 1 | 2.0 | 20 |
| GEO/05-Applied Geology | 165 | 0 | 0.0 | 0 |

*Table 6: List of SDSs with no Italian professors authoring any HCAs$_{(1\%)}$ in 2006-2010*

| SSD | UDA* | Research staff |
|---|---|---|
| ING-IND/13-Applied Mechanics for Machinery | IIE | 189 |
| GEO/05-Applied Geology | EAR | 165 |
| M-PSI/05-Social Psychology | ECS | 162 |
| ING-INF/07-Electric and Electronic Measurement Systems | IIE | 122 |
| ICAR/04-Road, Railway and Airport Construction | CEN | 108 |
| ICAR/05-Transport | CEN | 101 |
| VET/09-Clinical Veterinary Surgery | AVS | 93 |
| MED/19-Plastic Surgery | MED | 90 |
| ING-IND/15-Design and Methods for Industrial Engineering | IIE | 87 |
| ING-IND/09-Energy and Environmental Systems | IIE | 84 |
| AGR/10-Rural Construction and Environmental Land Management | AVS | 72 |
| MED/20-Pediatric and Infant Surgery | MED | 67 |
| VET/10-Clinical Veterinary Obstetrics and Gynaecology | AVS | 65 |
| AGR/20-Animal Husbandry | AVS | 53 |
| ING-IND/19-Nuclear Plants | IIE | 46 |
| ING-IND/12-Mechanical and Thermal Measuring Systems | IIE | 45 |
| MED/02-History of Medicine | MED | 35 |
| ING-IND/05-Aerospace Systems | IIE | 31 |
| AGR/14-Pedology | AVS | 30 |
| ING-IND/07-Aerospatial Propulsion | IIE | 30 |
| ING-IND/28-Excavation Engineering and Safety | IIE | 26 |
| ING-IND/02-Naval and Marine construction and installation | IIE | 20 |
| ING-IND/29-Raw Materials Engineering | IIE | 17 |
| AGR/06-Wood Technology and Woodland Management | AVS | 16 |
| ING-IND/20-Nuclear Measurement Tools | IIE | 15 |
| ING-IND/30-Hydrocarburants and Fluids of the Subsoil | IIE | 14 |
| ING-IND/18-Nuclear Reactor Physics | IIE | 13 |

\* MAT=Mathematics and computer sciences; PHY=Physics; CHE=Chemistry; EAR=Earth sciences; BIO=Biology; MED=Medicine; AVS=Agricultural and veterinary sciences; CEN=Civil engineering; IIE=Industrial and information engineering; PPS=Pedagogy and psychology; ECS=Economics and statistics

At the opposite extreme there are 44 SDSs where more than 10% of the total research staff are TSs$_{(1\%)}$. Table 7 presents the details for the top 20: half of the occurrences are in SDSs of Medicine, although heading the list are two Physics SDSs. In FIS/04, 39.5% of total research staff authored at least one HCA$_{(1\%)}$ over the five years examined, and in FIS/01 this percentage is 34.1. In the "over 30%" group we also find MED/15 and another Physics SDS, FIS/05. The top seven ranking are dominated by Physics and Medicine SDSs, with Medicine featuring twice again (MED/03, MED/12). Following the top seven SDSs are ING-IND/27 and VET/06, both with a percentage of TSs$_{(1\%)}$ out of national staff at around 20%. The first Chemistry SDS is CHIM/01, at 14$^{th}$ place on the list for incidence of TSs$_{(1\%)}$, with 16.1% of total research staff. The first and only Biology SDS on the list is BIO/11, with 15.4% TSs$_{(1\%)}$. We observe that the table of top 20 SDSs does not include any fields from Mathematics and computer sciences, Earth sciences, Civil engineering, Pedagogy and psychology or Economics and statistics. As a further test we carry out the same analysis as for Table 8 but extending the dataset of the HCAs to the top-5% cited articles. Comparing to the SDSs listed in Table 7 there are only six new entries, specifically VET/07, FIS/03, MED/10, M-PSI/02, CHIM/04 and CHIM/03. In general, correlating the ranking lists by TSs$_{(1\%)}$ and TSs$_{(5\%)}$, we obtain a Spearman correlation coefficient of 0.87 (two tail p-value=0.0000), confirming once again that at the level of fields, the threshold for identification of HCAs has little impact on the results of the analysis.

*Table 7: Top 20 SDSs with the highest percentage of Italian professors authoring HCAs$_{(1\%)}$ (full counting method; data 2006-2010)*

| SDS | UDA* | Research staff | TSs$_{(1\%)}$ | Incidence (%) |
|---|---|---|---|---|
| FIS/04-Nuclear and Subnuclear Physics | PHY | 162 | 64 | 39.5 |
| FIS/01-Experimental Physics | PHY | 1,000 | 341 | 34.1 |
| MED/15-Blood Diseases | MED | 173 | 53 | 30.6 |
| FIS/05-Astronomy and Astrophysics | PHY | 186 | 56 | 30.1 |
| MED/12-Gastroenterology | MED | 182 | 41 | 22.5 |
| MED/03-Medical Genetics | MED | 143 | 29 | 20.3 |
| ING-IND/27-Industrial and Technological Chemistry | IIE | 70 | 14 | 20.0 |
| VET/06-Parasitology and Parasitic Animal Diseases | AVS | 71 | 14 | 19.7 |
| MED/13-Endocrinology | MED | 256 | 46 | 18.0 |
| MED/01-Medical Statistics | MED | 106 | 19 | 17.9 |
| MED/08-Pathological Anatomy | MED | 324 | 58 | 17.9 |
| MED/06-Medical Oncology | MED | 132 | 23 | 17.4 |
| MED/11-Cardiovascular Diseases | MED | 272 | 45 | 16.5 |
| CHIM/01-Analytical Chemistry | CHE | 292 | 47 | 16.1 |
| FIS/02-Theoretical Physics, Mathematical Models and Methods | PHY | 361 | 58 | 16.1 |
| CHIM/12-Environmental Chem. and Chem. for cultural heritage | CHE | 69 | 11 | 15.9 |
| BIO/11-Molecular Biology | BIO | 221 | 34 | 15.4 |
| MED/09-Internal Medicine | MED | 1,092 | 161 | 14.7 |
| ING-IND/32-Electrical Convertors, Machines and Switches | IIE | 116 | 17 | 14.7 |
| MED/26-Neurology | MED | 426 | 60 | 14.1 |

* MAT=Mathematics and computer sciences; PHY=Physics; CHE=Chemistry; EAR=Earth sciences; BIO=Biology; MED=Medicine; AVS=Agricultural and veterinary sciences; CEN=Civil engineering; IIE=Industrial and information engineering; PPS=Pedagogy and psychology; ECS=Economics and statistics

*Table 8: First 20 SDS with the highest percentage of Italian professors authoring HCAs$_{(5\%)}$ (full counting method; data 2006-2010)*

| SDS | UDA* | Research staff | TSs$_{(5\%)}$ | Incidence (%) |
|---|---|---|---|---|
| FIS/04-Nuclear and Subnuclear Physics | PHY | 162 | 100 | 61.7 |
| MED/15-Blood Diseases | MED | 173 | 100 | 57.8 |
| FIS/01-Experimental Physics | PHY | 1,000 | 535 | 53.5 |
| MED/03-Medical Genetics | MED | 143 | 70 | 49.0 |
| FIS/05-Astronomy and Astrophysics | PHY | 186 | 89 | 47.8 |
| MED/06-Medical Oncology | MED | 132 | 62 | 47.0 |
| MED/01-Medical Statistics | MED | 106 | 48 | 45.3 |
| MED/12-Gastroenterology | MED | 182 | 80 | 44.0 |
| MED/13-Endocrinology | MED | 256 | 111 | 43.4 |
| ING-IND/27-Industrial and Technological Chemistry | IIE | 70 | 30 | 42.9 |
| VET/07-Veterinary Pharmacology and Toxicology | AVS | 47 | 20 | 42.6 |
| VET/06-Parasitology and Parasitic Animal Diseases | AVS | 71 | 30 | 42.3 |
| MED/08-Pathological Anatomy | MED | 324 | 136 | 42.0 |
| FIS/03-Material Physics | PHY | 458 | 189 | 41.3 |
| MED/10-Respiratory Diseases | MED | 131 | 54 | 41.2 |
| CHIM/01-Analytical Chemistry | CHE | 292 | 118 | 40.4 |
| M-PSI/02-Psychobiology and Physiological Psychology | PPS | 109 | 44 | 40.4 |
| CHIM/04-Industrial Chemistry | CHE | 150 | 60 | 40.0 |
| MED/26-Neurology | MED | 426 | 169 | 39.7 |
| CHIM/03-General and Inorganic Chemistry | CHE | 625 | 246 | 39.4 |

* MAT=Mathematics and computer sciences; PHY=Physics; CHE=Chemistry; EAR=Earth sciences; BIO=Biology; MED=Medicine; AVS=Agricultural and veterinary sciences; CEN=Civil engineering; IIE=Industrial and information engineering; PPS=Pedagogy and psychology; ECS=Economics and statistics

## 6. Discussion and conclusions

In order to formulate strategic goals, research policies should be based on sound analysis of the nation's research infrastructure. One of the important aspects of such analysis is the identification of the strengths and weaknesses in the various research fields. The results of such evaluations can be correlated to those from industrial analysis, in order to better align public research strengths and weaknesses in relation to those of industrial sectors and formulate priorities of intervention.

However the actual identification of national strengths and weaknesses at the level of research fields is a very challenging process, which scholars have only recently addressed, benefitting in part from current increases in the availability of bibliometric data. Still, the approaches employed to date are subject to critical flaws in regards to several purposes. Comparative studies between nations have resorted to the measure of the share of a country's articles, citations, or highly-cited articles relative to the world total. Such approaches generate size-dependent rankings in which the USA invariably results as the top nation in almost all scientific sectors. An obvious question is if these indicators permit us to affirm with certainty that the scientists of the top-rated countries are truly better than their colleagues in the rest of the world, or if the observations are more the effect of the absolute value of resources invested (an aspect where the USA is famously a leader among nations). The authors hold that a reliable comparative evaluation of research performance at field level must be conducted through the measure of total factor productivity. Unfortunately the data on input per scientific field in the various nations are not readily available, nor are those for output per research staff in the single fields. However the strategic analysis of national research systems, aimed at identifying strengths and weaknesses at field level, does not necessarily require comparison of productivity with other nations, although this would be of exceptional interest. In general, the useful objective is not so much to establish the national position in a given field compared to other countries, but rather to compare between fields within the nation. A pertinent question could be whether Italian physicists perform better than Italian mathematicians, and if among physicists it is the astrophysicists that are currently stronger than the theoretical physicists. The identification of such strengths and weaknesses can then inform research policies, development programs and allocation of funds. To determine whether we can indeed compare fields within a nation, we formulate an alternative approach for the measure of productivity, which although it should be improved to control for varying intensity of publication across fields, still permits control for size of input, and which above all is feasible. To apply the proposed method we take the Italian academic system as reference, which seems unique in offering the classification of each professor in one and only one research field, and we then reconstruct the scientific portfolio for each professor. Lacking availability of similar data for other nations for comparison of research productivity (although preparation of similar data seems possible), we resort to a second-best option, in which we compare research fields based on highly-cited articles. The measure of performance in a field is approximated as the fraction of national scientists working in that field who author highly-cited articles. The application of the method provides a strengths and weaknesses analysis in fields of comparable intensity of publication of the 200 fields investigated, and should certainly prove interesting to the policy maker. Variations conducted to control for the effect of differing intensity of co-authorships between fields, and to observe the results from differing the threshold used to identify the HCAs, indicate substantial robustness in the method. Still another possibility would be approximate the relative performance of the nation's scientific fields through a weighted combination of $TS_{(1\%)}$, $TS_{(5\%)}$ and the like. Future research should focus on overcoming the limitation due to the varying intensity of publication across fields.

**References**

Abramo, G., D'Angelo, C.A. (2007). Measuring science: Irresistible temptations, easy shortcuts and dangerous consequences. *Current Science*, 93(6), 762-766.

Abramo, G., D'Angelo, C.A., Cicero, T. (2012). What is the appropriate length of the publication period over which to assess research performance? *Scientometrics*, 93(3), 1005-1017.

Abramo, G., D'Angelo, C.A., Murgia, G. (2013). The collaboration behaviors of scientists in Italy: a field level analysis. *Journal of Informetrics*, 7(2), 442-454.

Adams, J. (1998). Benchmarking international research. *Nature*, 396, 615–618.

Adams, J. (1990). Fundamental stocks of knowledge and productivity growth. *Journal of Political Economy*, 98(4), 673–702.

Boaz, A., Ashby, D. (2003). *Fit for purpose? Assessing research quality for evidence based policy and practice*. Retrieved from ESRC UK Centre for Evidence Based Policy and Practice, http://www.kcl.ac.uk/sspp/departments/politicaleconomy/research/cep/pubs/papers/assets/wp11.pdf, last accessed on June 23, 2014.

Bornmann, L., Leydesdorff, L. (2011). Which cities produce more excellent papers than can be expected? A new mapping approach—using Google Maps—based on statistical significance testing. *Journal of the American Society of Information Science and Technology*, 62(10), 1954-1962.

Bornmann, L., Leydesdorff, L., Walch-Solimena, C., Ettl, C. (2011). Mapping excellence in the geography of science: an approach based on Scopus data. Journal of Informetrics, 5(4), 537-546.

D'Angelo, C.A., Giuffrida, C., Abramo, G. (2011). A heuristic approach to author name disambiguation in bibliometrics databases for large-scale research assessments. *Journal of the American Society for Information Science and Technology*, 62(2), 257–269.

Etzkowitz, H., Webster, A., Gebhardt, C., Cantisano Terra, B.R. (2000). The future of the university and the university of the future: evolution of ivory tower to entrepreneurial paradigm. *Research Policy*, 29(2), 313-330.

Glänzel, W., Schubert, A. (1988). Characteristic scores and scales in assessing citation impact. *Journal of Information Science*, 14(2), 123-127

Grant, J., Brutscher, P.C., Kirk, S. E., Butler, L., Wooding, S. (2010). *Capturing Research Impacts: A review of international practice*. Cambridge, UK: Rand Europe.

Griliches, Z. (1998). *R&D and Productivity*. Chicago University Press, Chicago. ISBN 9780226308869

Hauser, J.R., Zettelmeyer, F. (1997): Metric to evaluate R, D & E. *Research Technology Management,* 40(4), 32-38.

Henderson, R., Jaffe, A., Trajtenberg, M. (1998). Universities as a source of commercial technology: a detailed analysis of university patenting, 1965-1988. *Review of*

*Economics and Statistics*, 65, 119-127.

Hicks, D. (2012). Performance-based university research funding systems. *Research Policy*, 41(2), 251-261.

King, D.A. (2004). The scientific impact of nations - What different countries get for their research spending. *Nature*, 430, 311–316.

Mansfield, E. (1991). Academic Research and Industrial Innovation. *Research Policy*, 20(1), 1–12.

Mansfield, E. (1995). Academic research underlying industrial innovations: sources, characteristics, and financing. *Review of Economics and Statistics*, 77, 55–65.

Martin, B.R., Irvine, J. (1983). Assessing basic research: some partial indicators of scientific progress in radio astronomy. *Research Policy*, 12(2), 61-90.

Martin, B.R., Salter, A., Hicks, D., Pavitt, K., Senker, J., Sharp, M., Von Tunzelmann, N. (1996). *The Relationship Between Publicly Funded Basic Research and Economic Performance: A SPRU Review*. HM Treasury, London.

May, R.M. (1997). The scientific wealth of nations. *Science,* 275(5301), 793-796.

OECD-Organisation for Economic Co-operation and Development (1997). The evaluation of scientific research: Selected experiences. Paris: OECD.

Pislyakov, V., Shukshina, E. (2012). Measuring Excellence in Russia: Highly Cited Papers, Leading Institutions, Patterns of National and International Collaboration. *17th International Conference on Science and Technology Indicators (STI)*, 5-8 September, 2012 in Montreal, Quebec, Canada.

Rosenberg, N., Nelson, R. (1994). American universities and technical advance in industry. *Research Policy*, 23, 323–348.

Tijssen, R.J.W. (2003). Scoreboards of research excellence. *Research Evaluation*, 12 (2), 91-103.

Tijssen, R.J.W., Visser, M.S., Van Leeuwen, T.N. (2002). Benchmarking international scientific excellence: Are highly cited research papers an appropriate frame of reference? *Scientometrics,* 54(3), 381-97.

Werner, B.M., Souder, W.E. (1997): Measuring R&D performance – state of the art. *Research Technology Management,* 40(2), 34-42.